# *J*-matrix method of scattering for inverse-square singular potentials with supercritical coupling II. Regularization

Abdulaziz D. Alhaidari[(a)†], Hocine Bahlouli[(b)], S. M. Al-Marzoug[(b)], and Carlos P. Aparicio[(c)]

[(a)] *Saudi Center for Theoretical Physics, P. O. Box 32741, Jeddah 21438, Saudi Arabia*

[(b)] *Physics Department, King Fahd University of Petroleum & Minerals, Dhahran 31261, Saudi Arabia*

[(c)] *Alameda de San Antón 45 3ºI, CP 30205 Cartagena, Murcia, Spain*

**Abstract**: This paper is a continuation of the previous one [Journal **xx**, xxxxx (2022)]. Here, we reformulate the same *J*-matrix theory by regularizing the inverse square singular potential. The objective is to restore rapid convergence of the calculation in the theory and recover the conventional tridiagonal representation. Partial success is achieved.

## I. Introduction

In our previous paper [1], which will be referred to as Paper I from this point onward, we were able to extend the *J*-matrix method of scattering to handle inverse square singular potentials with supercritical coupling. We were able to develop a consistent theory without the need for regularization, which is typically needed in such situations. However, we had to pay a price that was exhibited in two ways:

(1) Slow convergence of the computational algorithm, and

(2) We were obliged to work in a larger representation space.

The first is a major disadvantage since one of the prime features of the conventional *J*-matrix method is the rapid convergence in calculating the scattering matrix. The second, which is less serious, forced us to deal with objects that satisfy five-term recursion relations instead of the usual three-term recursion relations satisfied by orthogonal polynomials. In this second paper, we present an alternative version of the same theory where we perform a regularization of the inverse-squared singular potential. The objective is an attempt to eliminate one or, hopefully, both above-mentioned disadvantages of the un-regularized theory in Paper I. Some examples of the regularization strategies that were adopted in the literature include [2-8], but not limited to

- Infinite square barrier regularization: Introducing an ultraviolet cutoff by inserting an infinite square barrier of small width $r_0$ at the origin (i.e., inserting a small hard sphere centered at the origin) effectively limiting configuration space to $r \geq r_0$ and removing the poison-tooth from the potential function at the origin,

- Finite square well regularization: Introducing a potential cutoff of size $r_0$ by replacing the potential near the origin (for $r \leq r_0$) with a constant (i.e., inserting a small soft sphere centered at the origin),

---

[†] Corresponding Author. Email: haidari@sctp.org.sa

- Delta shell regularization: Inserting a delta potential function $\delta(r-r_0)$ around the origin and very close to it, etc.

However, the regularization scheme that we adopt here follows from a strategy that we have introduced in Ref. [9] where we maintain the same type of singularity at the origin but we tame it. That is, we regularize the inverse square singular potential by rewriting it in the following two-parameter potential form

$$V(r) = \begin{cases} -\dfrac{A_0/2}{r^2} & , r \leq r_0 \\ -\dfrac{A/2}{r^2} + U(r) & , r > r_0 \end{cases} \tag{1}$$

where $U(r)$ is the short range non-singular component of the potential and $r_0$ is a very small adjustable radial parameter such that $\lambda r_0 \ll 1$ with $\lambda$ being some appropriate scale parameter of inverse length dimension. The second parameter is the regularized potential strength parameter $A_0$. Super-criticality of the original potential requires that $(\ell+\tfrac{1}{2})^2 < A$, where $\ell$ is the angular momentum quantum number. However, we tame the singularity at the origin by requiring that $(\ell+\tfrac{1}{2})^2 > A_0$ making the coupling of the inverse square potential near the origin subcritical [9]. This allows us to treat the reference problem near the origin ($0 \leq r \leq r_0$) in a manner similar to that in Ref. [10]. Additionally, we could also impose (if possible) continuity of the potential at $r = r_0$ giving $A_0 = A - 2r_0^2 U(r_0)$. This leaves us with only one regularization parameter, $r_0$. The sub-criticality constraint requires that the choice of $r_0$ must be made such that $2r_0^2 U(r_0) > A - (\ell+\tfrac{1}{2})^2$. Therefore, $U(r_0)$ should be positive otherwise we cannot impose continuity of the potential at $r = r_0$, which is after all not necessary.

In the atomic units $\hbar = m = 1$, the time independent Schrödinger equation for a point particle in the field of a central potential $V(r)$ reads as follows

$$\left[-\frac{1}{2}\frac{d^2}{dr^2} + \frac{\ell(\ell+1)}{2r^2} + V(r)\right]\psi(r) = [H_0 + U(r)]\psi(r) = E\psi(r), \tag{2}$$

where the reference Hamiltonian in the J-matrix formalism is obtained using (1) as follows

$$H_0 = -\frac{1}{2}\frac{d^2}{dr^2} + \frac{\ell(\ell+1)}{2r^2} - \frac{1}{2r^2} \times \begin{cases} A_0 & , r \leq r_0 \\ A & , r > r_0 \end{cases} \tag{3}$$

For $r \leq r_0$, the solution of the reference problem is obtained in [10] as follows

$$\psi(r) = \sqrt{kr}\left[B_+ J_\nu(kr) + B_- J_{-\nu}(kr)\right], \tag{4}$$

where $k = \sqrt{2E}$, $\nu = \sqrt{(\ell+1/2)^2 - A_0}$, and $B_\pm$ are arbitrary normalization parameters. The Bessel function $J_\nu(kr)$ is regular everywhere whereas $J_{-\nu}(kr)$ diverges at the origin. On the other hand, for $r > r_0$ the solution of the reference problem is obtained in Paper I as

$$\psi(r) = \sqrt{kr}\left[A_+ H^+_{i\mu}(kr) + A_- H^-_{i\mu}(kr)\right], \tag{5}$$

where $\mu = \sqrt{A-(\ell+1/2)^2}$, $H_\alpha^\pm(z) = J_\alpha(z) \pm iY_\alpha(z)$, $Y_\alpha(z) = [\cos(\alpha\pi)J_\alpha(z) - J_{-\alpha}(z)]/\sin(\alpha\pi)$ and $A_\pm$ are arbitrary normalization constants. The asymptotics boundary conditions could be written as follows

$$\lim_{r\to\infty}\begin{Bmatrix}\psi_{reg}(r)\\ \psi_{irr}(r)\end{Bmatrix} = \sqrt{\frac{2}{\pi}}\begin{Bmatrix}\cos(kr+\vartheta-\frac{\pi}{4})\\ \sin(kr+\vartheta-\frac{\pi}{4})\end{Bmatrix} = \sqrt{\frac{2}{\pi}}\begin{Bmatrix}\frac{1}{2}\left[e^{+i(kr+\vartheta-\pi/4)}+e^{-i(kr+\vartheta-\pi/4)}\right]\\ \frac{1}{2i}\left[e^{+i(kr+\vartheta-\pi/4)}-e^{-i(kr+\vartheta-\pi/4)}\right]\end{Bmatrix} \quad (6)$$

where $\vartheta$ is the scattering phase shift for the regularized reference problem that depends on the energy, physical parameters $\{\ell,A\}$ and regularization parameters $\{r_0,A_0\}$. Using these boundary conditions and $\lim_{y\to\infty}\sqrt{y}H_{i\mu}^\pm(y) = \sqrt{\frac{2}{\pi}}e^{\pm\pi\mu/2}e^{\pm i(y-\pi/4)}$ gives either $A_\pm = \frac{1}{2}e^{\mp\pi\mu/2}e^{\pm i\vartheta}$ or $A_\pm = \frac{\pm 1}{2i}e^{\mp\pi\mu/2}e^{\pm i\vartheta}$. Thus, the regular and irregular reference wavefunctions become

$$\psi_{reg}(r) = \sqrt{kr}\begin{cases}J_\nu(kr) &, r \le r_0\\ \frac{1}{2}\left[e^{-\pi\mu/2}e^{i\vartheta}H_{i\mu}^+(kr) + e^{\pi\mu/2}e^{-i\vartheta}H_{i\mu}^-(kr)\right] &, r \ge r_0\end{cases} \quad (7a)$$

$$\psi_{irr}(r) = \sqrt{kr}\begin{cases}B_+J_\nu(kr) + B_-J_{-\nu}(kr) &, r \le r_0\\ \frac{1}{2i}\left[e^{-\pi\mu/2}e^{i\vartheta}H_{i\mu}^+(kr) - e^{\pi\mu/2}e^{-i\vartheta}H_{i\mu}^-(kr)\right] &, r \ge r_0\end{cases} \quad (7b)$$

Comparing these reference solutions to those obtained in Paper I, we see that the effect of regularization is in the introduction of the phase shift $e^{\pm i\vartheta}$ in the asymptotics of the reference wavefunction. This effect will of course appear in the scattering matrix that will be calculated in section III below. Matching the two reference wavefunctions (7) and their spatial derivatives at $r = r_0$ will determine $B_\pm$ and $\vartheta$. After some elementary but laborious calculations, we obtain the following $2\times 2$ matrix equation that determines $B_\pm$

$$\begin{pmatrix}M_{11} & M_{12}\\ M_{21} & M_{22}\end{pmatrix}\begin{pmatrix}B_+\\ B_-\end{pmatrix} = \begin{pmatrix}u_+\\ u_-\end{pmatrix}, \quad (8a)$$

where $M_{11} = J_\nu(kr_0)$, $M_{12} = J_{-\nu}(kr_0)$, $u_- = \left(\gamma_\mu^+ + \gamma_\mu^-\right)J_\nu(kr_0)$, and

$$u_+ = \frac{-1}{\gamma_\mu^+ + \gamma_\mu^-}\left[\lambda_\nu^+ J_{\nu-1}(kr_0) + \lambda_\nu^- J_{\nu+1}(kr_0) + i\left(\gamma_\mu^- - \gamma_\mu^+\right)J_\nu(kr_0)\right]. \quad (8b)$$

$$M_{21} = \lambda_\nu^+ J_{\nu-1}(kr_0) + \lambda_\nu^- J_{\nu+1}(kr_0) + i\left(\gamma_\mu^- - \gamma_\mu^+\right)J_\nu(kr_0), \quad (8c)$$

$$M_{22} = -\lambda_\nu^+ J_{-\nu+1}(kr_0) - \lambda_\nu^- J_{-\nu-1}(kr_0) + i\left(\gamma_\mu^- - \gamma_\mu^+\right)J_{-\nu}(kr_0). \quad (8d)$$

Where we have defined $\gamma_\mu^\pm := \frac{1}{2\xi_\mu^\pm}\left(\lambda_{i\mu}^+ \xi_{\mu+i}^\pm - \lambda_{i\mu}^- \xi_{\mu-i}^\pm\right)$, $\xi_\alpha^\pm := \frac{1}{2}e^{\mp\pi\alpha/2}H_{i\alpha}^\pm(kr_0)$, and $\lambda_\alpha^\pm := \frac{1}{2\alpha} \pm 1 = -\lambda_{-\alpha}^\mp$. Finally, the phase shift is obtained using $B_\pm$ found above as

$$e^{\pm i\vartheta} = \frac{\pm i}{2\xi_\mu^\pm}\left[(B_+ \mp i)J_\nu(kr_0) + B_-J_{-\nu}(kr_0)\right]. \quad (8e)$$

In these calculations, we used the following differential and recursion properties of the Bessel function [11]

$$\frac{d}{dx}J_\nu(x) = \frac{1}{2}[J_{\nu-1}(x) - J_{\nu+1}(x)], \tag{9a}$$

$$\frac{1}{x}J_\nu(x) = \frac{1}{2\nu}[J_{\nu-1}(x) + J_{\nu+1}(x)]. \tag{9b}$$

Giving

$$2\frac{d}{dx}\left[\sqrt{x}F_\nu(x)\right] = \sqrt{x}\left[\lambda_\nu^+ F_{\nu-1}(x) + \lambda_\nu^- F_{\nu+1}(x)\right], \tag{9c}$$

where $F_\nu(x)$ stands for either $J_\nu(x)$, $Y_\nu(x)$, or $H_\nu^\pm(x)$.

## II. *J*-matrix solution of the reference problem

Now, the *J*-matrix solution of the reference problem for $r \leq r_0$ (for $r \geq r_0$) are found in Ref. [10] (in Paper I), respectively. These are written as bounded convergent series of square-integrable functions with coefficients that satisfy three-term (five-term) recursion relations, respectively. In analogy with the splitting of configurations space for the potential (1), reference Hamiltonian (3) and reference wavefunction (7), we split the basis set as follows

$$\phi_n(r) = \begin{cases} \varphi_n(r), & r \leq r_0 \\ \chi_n(r), & r > r_0 \end{cases} = e^{-\lambda r/2} \begin{cases} \sqrt{\frac{\lambda \Gamma(n+1)}{\Gamma(n+2\nu+1)}}(\lambda r)^{\nu+\frac{1}{2}} L_n^{2\nu}(\lambda r), & r \leq r_0 \\ \sqrt{\frac{\lambda \Gamma(n+1)}{\Gamma(n+2\zeta+1)}}(\lambda r)^{\zeta+1} L_n^{2\zeta}(\lambda r), & r > r_0 \end{cases} \tag{10}$$

where $\zeta$ is a positive dimensionless parameter. Here, we consider a representation of the reference problem in the "Laguerre basis" only. Representation in the "oscillator basis" where the argument of the Laguerre polynomial is $\lambda^2 r^2$ will not be treated. From this point forward, we make the natural choice for the parameter $\zeta$ as $\zeta = \mu$. We start by writing the sine-like and cosine-like *J*-matrix solutions of the reference problem as follows

$$\psi_{\sin}(r) = \sum_{n=0}^{\infty} S_n(E)\phi_n(r) = \begin{cases} \sum_{n=0}^{\infty} S_n(E)\varphi_n(r), & r \leq r_0 \\ \sum_{n=0}^{\infty} \mathcal{S}_n(E)\chi_n(r), & r > r_0 \end{cases} = \psi_{reg}(r) \tag{11a}$$

$$\psi_{\cos}(r) = \sum_{n=0}^{\infty} C_n(E)\phi_n(r) = \begin{cases} \sum_{n=0}^{\infty} \mathcal{C}_n(E)\varphi_n(r), & r \leq r_0 \\ \sum_{n=0}^{\infty} C_n(E)\chi_n(r), & r > r_0 \end{cases}, \tag{11b}$$

where the expansion coefficients $\{S_n, \mathcal{S}_n, \mathcal{C}_n, C_n\}$ are to be determined shortly below. It should be noted that $\psi_{\sin}(r)$ must be identical to $\psi_{reg}(r)$ everywhere whereas $\psi_{\cos}(r)$, being regular, agrees with $\psi_{irr}(r)$ only asymptotically. Now, the expansion coefficients $\{S_n, \mathcal{C}_n\}$ satisfy the following three-term recursion relation [10]

$$\alpha_n P_n(E) + \beta_{n-1} P_{n-1}(E) + \beta_n P_{n+1}(E) = 0, \tag{12}$$

where $P_n$ stands for either $S_n$ or $C_n$ and

$$\alpha_n = (2n + 2\nu + 1)\cos\theta, \tag{13a}$$

$$\beta_n = -\sqrt{(n+1)(n+2\nu+1)}, \tag{13b}$$

where $\cos\theta = \frac{4\sigma^2-1}{4\sigma^2+1}$, $\sigma = k/\lambda$, and $0 < \theta \leq \pi$. The recursion relation (12) is valid for $n \geq 1$ and it is solved for $\{S_n\}_{n=2}^\infty$ and $\{C_n\}_{n=2}^\infty$ starting with the following initial values [10]

$$S_0(E) = \frac{\Gamma(\nu+\tfrac{1}{2})}{\sqrt{2\pi\lambda\Gamma(2\nu+1)}}(2\sin\theta)^{\nu+\tfrac{1}{2}}, \tag{14a}$$

$$S_1(E) = \frac{2\Gamma(\nu+\tfrac{3}{2})(\cos\theta)}{\sqrt{2\pi\lambda\Gamma(2\nu+2)}}(2\sin\theta)^{\nu+\tfrac{1}{2}}, \tag{14b}$$

$$C_0(E) = \frac{2\Gamma(\nu+1)}{\sqrt{\pi}\Gamma(\nu+\tfrac{1}{2})}\tau(E)S_0(E), \tag{15a}$$

$$C_1(E) = \frac{2}{\sqrt{\pi}}\Gamma(\nu+1)\left[\frac{\tau(E)S_1(E)}{\Gamma(\nu+\tfrac{1}{2})} - \eta\frac{2^\nu(\sin\theta)^{\tfrac{1}{2}-\nu}}{\sqrt{\pi\lambda\Gamma(2\nu+2)}}\right], \tag{15b}$$

where $\tau(E) = (\cos\theta)\,_2F_1\!\left(\tfrac{1}{2},\nu+1;\tfrac{3}{2};\cos^2\theta\right)$ and $\eta$ is a parameter obtained by matching the two series of $\psi_{\cos}(r)$ at $r = r_0$ [12]. On the other hand, the expansion coefficients $\{S_n, C_n\}$ satisfy the following five-term recursion relation given in Paper I

$$a_n P_n(E) + b_{n-1} P_{n-1}(E) + b_n P_{n+1}(E) + c_{n-2} P_{n-2}(E) + c_n P_{n+2}(E) = 0, \tag{16}$$

where $P_n$ stands for either $S_n$ or $C_n$ and

$$a_n = \frac{8\mu^2 - 1}{4\sigma^2 + 1} + (\cos\theta)(2n+2\mu+1)^2 + \left[2n^2 + (2n+1)(2\mu+1)\right], \tag{17a}$$

$$b_n = -4\sigma(\sin\theta)(n+\mu+1)\sqrt{(n+1)(n+2\mu+1)}, \tag{17b}$$

$$c_n = \sqrt{(n+1)(n+2)(n+2\mu+1)(n+2\mu+2)}, \tag{17c}$$

The recursion relation (16) is valid for $n \geq 2$ and it is solved for $\{S_n\}_{n=4}^\infty$ and $\{C_n\}_{n=4}^\infty$ starting with the initial values $\{S_n\}_{n=0}^{n=3}$ and $\{C_n\}_{n=0}^{n=3}$. If we define $\mathcal{F}_n^\pm(E) = S_n(E) \pm iC_n(E)$, then we can compute $\{\mathcal{F}_n^\pm\}_{n=0}^\infty$ as shown in Paper I starting with $\mathcal{F}_0^\pm$ and $\mathcal{F}_1^\pm$ only because $\mathcal{F}_2^\pm$ and $\mathcal{F}_3^\pm$ are obtained from these using Eq. (19) therein. Following exactly the same procedure as in Paper I, we obtain the following expressions for $\mathcal{F}_0^\pm(E)$ and $\mathcal{F}_1^\pm(E)$

$$\mathcal{F}_0^\pm(E) = \frac{\pm e^{\pm i\vartheta}}{\sinh(\mu\pi)} \sqrt{\frac{\sigma}{\Gamma(2\mu+1)}} \left(\sqrt{4\sigma^2+1}\right)^{-\mu-\frac{1}{2}}$$
$$\left[ e^{\pm\mu\pi/2} P_{\mu-\frac{1}{2}}^{-i\mu}\left(1/\sqrt{4\sigma^2+1}\right) - e^{\mp\mu\pi/2} P_{\mu-\frac{1}{2}}^{i\mu}\left(1/\sqrt{4\sigma^2+1}\right) \right]$$
(18a)

$$\mathcal{F}_1^\pm(E) = \sqrt{2\mu+1}\, \mathcal{F}_0^\pm(E) + \frac{\mp e^{\pm i\vartheta}}{\sinh(\mu\pi)} \sqrt{\frac{\sigma}{\Gamma(2\mu+2)}} \left(\sqrt{4\sigma^2+1}\right)^{-\mu-\frac{3}{2}}$$
$$\left[ e^{\pm\mu\pi/2} P_{\mu+\frac{1}{2}}^{-i\mu}\left(1/\sqrt{4\sigma^2+1}\right) - e^{\mp\mu\pi/2} P_{\mu+\frac{1}{2}}^{i\mu}\left(1/\sqrt{4\sigma^2+1}\right) \right]$$
(18b)

where $P_\gamma^\lambda(x)$ is the associated Legendre function of the first kind defined by E. (16) in Paper I. One should note the effect of regularization in the presence of the phase shift factor $e^{\pm i\vartheta(E)}$. For a given set of physical parameters $\{\ell, A\}$ and a set of regularization parameters $\{r_0, A_0\}$, Figure 1 shows $\psi_{\sin}(r)$ and $\psi_{\cos}(r)$ superimposed on $\psi_{reg}(r)$ and $\psi_{irr}(r)$, respectively. It is evident that $\psi_{\sin}(r)$ agrees very well with $\psi_{reg}(r)$ for all $r$ and that the agreement improves with increasing the basis size $N$. However, $\psi_{\cos}(r)$ and $\psi_{irr}(r)$ agree very well only for $r \geq r_0$.

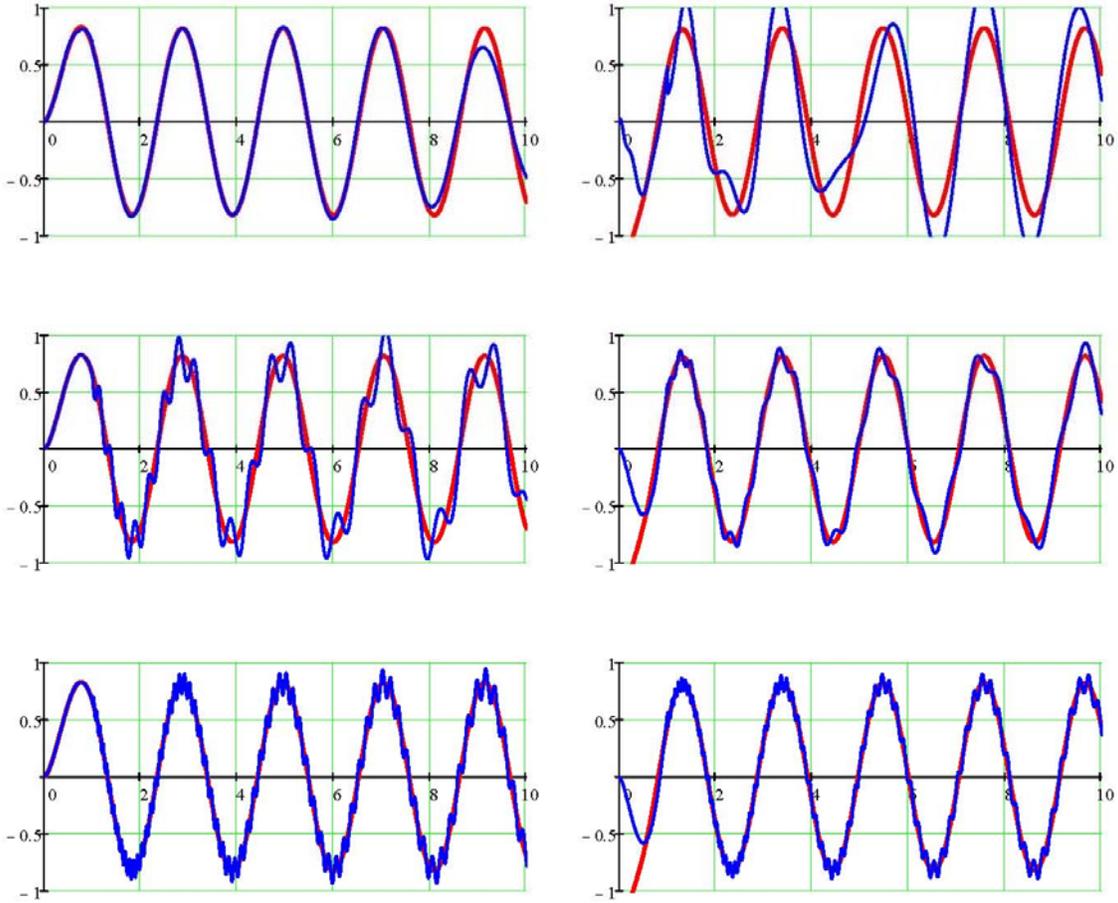

**FIG. 1**: Plots of the reference solution as given by Eq. (7) in red superimposed on the $J$-matrix solution as given by Eq. (11) in blue. The sum in Eq. (11) is truncated up to $n = N$ and we took $\ell = 1$, $A = 3$, and the energy parameter $\sigma = 3$. The regularization parameters were taken as $A_0 = 1$ and $\lambda r_0 = 1$ (this large $\lambda r_0$ value was chosen to enhance the visual presentation in the figure). The plots on the left column are for $\psi_{reg}(r)$ and $\psi_{\sin}(r)$ whereas those on the right are for $\psi_{irr}(r)$ and $\psi_{\cos}(r)$. The three sizes of the basis set are taken as $N = 100, 1000, 10000$ from top to bottom row. The horizontal axis is the radial coordinate $r$ in units of $\lambda^{-1}$.

Figure 2 is a plot of the regular solution $\psi_{reg}(r)$ very close to the origin superimposed on $\psi_{\sin}(r)$ for a basis size $N = 10000$. Comparing it to Figure 2 in Paper I, with the same basis size, it becomes evident that convergence has greatly improved. When comparing figures, one should make note of the magnification scale of the vertical axis. For example, the scale in the top figure here is about 5 times that of the corresponding figure in Paper I, whereas the scale in the bottom figure here is about 50 times that of the corresponding figure in Paper I.

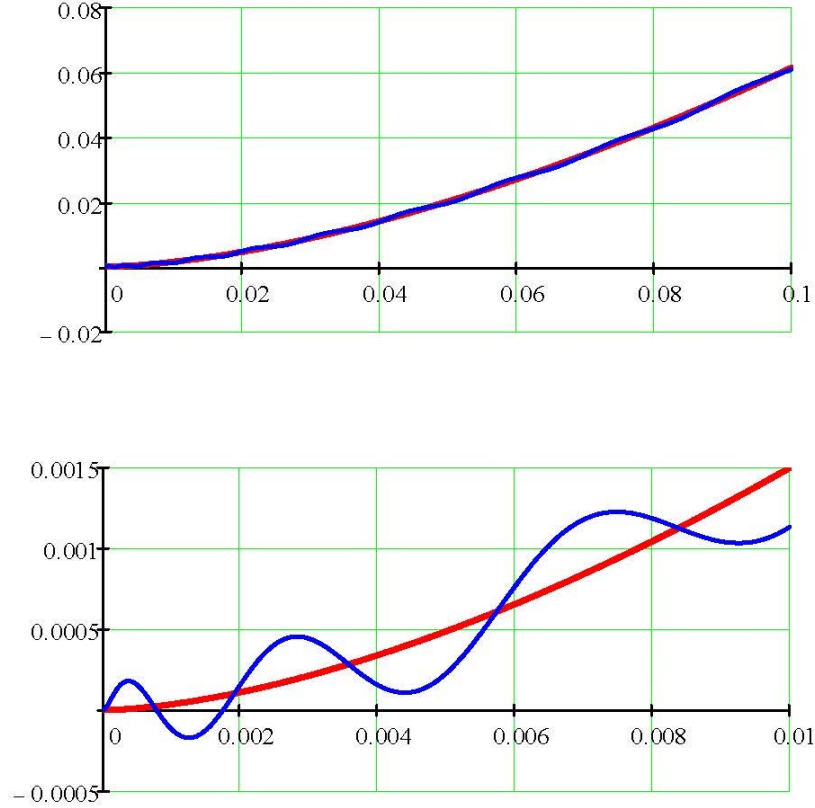

**FIG. 2**: Plot of the reference solution $\psi_{reg}(r)$ in red superimposed on the regularized $J$-matrix solution $\psi_{\sin}(r)$ in blue very close to the origin and for basis size $N = 10000$

## III. The scattering matrix

We follow exactly the same procedure as shown in Paper I to calculate the scattering matrix. Thus, we obtain

$$S(E) = T_{N-1} \frac{1 + \left(G_{N-1,N-1} \mathcal{J}_{N-1,N} + G_{N-1,N-2} \mathcal{J}_{N-2,N}\right) R_N^+ + G_{N-1,N-1} \mathcal{J}_{N-1,N+1} R_{N+1}^+ R_N^+}{1 + \left(G_{N-1,N-1} \mathcal{J}_{N-1,N} + G_{N-1,N-2} \mathcal{J}_{N-2,N}\right) R_N^- + G_{N-1,N-1} \mathcal{J}_{N-1,N+1} R_{N+1}^- R_N^-}, \quad (19)$$

where $N$ is some large enough integer. In fact, we should have expected that the $S$-matrix, being an asymptotic object far disconnected from the regularization region, to have the exact same mathematical form. However, remanence of the regularization procedure will persist within its

–7–

elements. For example, in the definition of $T_n(E)$ and $R_n^\pm(E)$, we use the expansion coefficients $\{\mathcal{F}_n^\pm(E)\}_{n=0}^\infty$ that depend on the regularization parameters $\{r_0, A_0\}$ via the phase shift $e^{\pm i\vartheta}$ not those in Paper I in terms of $\{F_n^\pm(E)\}_{n=0}^\infty$. Therefore, we write

$$T_n(E) = \frac{\mathcal{F}_n^+(E)}{\mathcal{F}_n^-(E)} = \frac{\mathcal{S}_n(k) + i\mathcal{C}_n(k)}{\mathcal{S}_n(k) - i\mathcal{C}_n(k)}, \quad R_n^\pm(E) = \frac{\mathcal{F}_n^\pm(E)}{\mathcal{F}_{n-1}^\pm(E)} = \frac{\mathcal{S}_n(k) \pm i\mathcal{C}_n(k)}{\mathcal{S}_{n-1}(k) \pm i\mathcal{C}_{n-1}(k)}. \tag{20}$$

On the other hand, the matrix elements of the reference wave operator $\mathcal{J}_{N-1,N}(E)$, $\mathcal{J}_{N-1,N+1}(E)$ and $\mathcal{J}_{N-2,N}(E)$ are identical to those in Paper I reading

$$\frac{-2}{\lambda^2}\mathcal{J}_{n,m}(E) = \left(\sigma^2 + \tfrac{1}{4}\right)\left(a_n \delta_{n,m} + b_{n-1}\delta_{n,m+1} + b_n \delta_{n,m-1} + c_{n-2}\delta_{n,m+2} + c_n \delta_{n,m-2}\right), \tag{21}$$

for $n,m \gg 1$. Moreover, the finite $N \times N$ submatrix block on the top-left corner of Eq. (22) in Paper I becomes the representation of the wave operator $H_0 + U - E$ in a suitable finite basis set, which we take as $\{\chi_n\}_{n=0}^{N-1}$ and whose size $N$ must be large enough to give a faithful representation of the short range potential $U(r)$. This block is used to calculate the elements of the finite Green's function $G_{N-1,N-1}(E)$ and $G_{N-1,N-2}(E)$ as explained in Appendix C of Paper I.

## IV. Conclusion

In this paper, which is the second following Paper I [1], we presented an alternative version of the same $J$-matrix theory of quantum scattering for inverse-square singular potentials with supercritical coupling. Here, we perform a regularization of the inverse square potential in an attempt to restore rapid convergence and recover the tridiagonal structure of the theory. However, contrary to most traditional regularization schemes, we kept the same type of singularity of the potential in the regularization region but changed its strength to become subcritical.

We were partially successful on both counts. Convergence has improved, especially for higher energies, but not very much. The tridiagonal structure has been recovered but not in the asymptotic region where it counts for $S$-matrix calculation.

## Acknowledgements

This work is dedicated to the memory of our departed friend, colleague and collaborator, Mohammed S. Abdelmonem.

## References

[1] A. D. Alhaidari, H. Bahlouli, C. P. Aparicio, and S. M. Al-Marzoug, *J-matrix method of scattering for inverse-square singular potentials with supercritical coupling* I. *No regularization*, Journal **xx**, xxxxxx (2022)

[2]  K. S. Gupta and S. G. Rajeev, *Renormalization in quantum mechanics*, Phys. Rev. D **48**, 5940 (1993)

[3]  H. E. Camblong, L. N. Epele, H. Fanchiotti, and C. A. García Canal, *Renormalization of the Inverse Square Potential*, Phys. Rev. Lett. **85**, 1590 (2000)

[4]  M. Bawin and S. A. Coon, *Singular inverse square potential, limit cycles, and self-adjoint extensions*, Phys. Rev. A **67**, 042712 (2003)

[5]  E. Braaten and D. Phillips, *Renormalization-group limit cycle for the $1/r^2$ potential*, Phys. Rev. A **70**, 052111 (2004)

[6]  A. M. Essin and D. J. Griffiths, *Quantum mechanics of the $1/x^2$ potential*, Am. J. Phys. **74**, 109 (2006)

[7]  D. Bouaziz and M. Bawin, *Regularization of the singular inverse square potential in quantum mechanics with a minimal length*, Phys. Rev. A **76**, 032112 (2007)

[8]  D. Bouaziz and M. Bawin, *Singular inverse-square potential: Renormalization and self-adjoint extensions for medium to weak coupling*, Phys. Rev. A **89**, 022113 (2014)

[9]  A. D. Alhaidari, *Renormalization of the strongly attractive inverse square potential: Taming the singularity*, Found. Phys. **44**, 1049 (2014)

[10] A.D. Alhaidari, H. Bahlouli, S. M. Al-Marzoug and M.S. Abdelmonem, *J-matrix method of scattering for potentials with inverse square singularity: The real representation*, Phys. Scr. **90**, 055205 (2015)

[11] See, for example, Sec. 3.1.1 in: W. Magnus, F. Oberhettinger, and R. P. Soni, *Formulas and Theorems for the Special Functions of Mathematical Physics*, 3rd ed. (Springer, Berlin, 1966)

[12] The parameter $\eta$ depends on the energy, angular momentum and the regularization parameters. However, it is independent of the inverse singularity strength *A*.